%% file: InSilicoPlants_Sub_01.tex
\documentclass[unnumsec,webpdf,contemporary,large]{oup-authoring-template}%

\usepackage[utf8]{inputenc}
\usepackage{color}
\usepackage{verbatim}
\usepackage{units}
\usepackage{amsmath}
\usepackage{amssymb}
\usepackage{graphicx}

\usepackage{bm}
\usepackage{amsfonts}
\usepackage{comment}
\usepackage{subfig}

\theoremstyle{thmstyleone}%
\theoremstyle{thmstyletwo}%
\theoremstyle{thmstylethree}%

\newcommand{\T}{^{\mathsf{T}}}
\newcommand{\cne}[2]{#1 \, \text{x} \, 10^{#2}}
\newcommand{\cn}[2]{$#1 \, \text{x} \, 10^{#2}$}

\begin{document}

\journaltitle{Journal Title Here}
\DOI{DOI HERE}
\copyrightyear{2022}
\pubyear{2019}
\access{Advance Access Publication Date: Day Month Year}
\appnotes{Paper}

\firstpage{1}


\title{Data-Driven Modeling of Photosynthesis Regulation Under Oscillating Light Condition - Part I: In-Silico Exploration}

\author[1]{Christian Portilla}
\author[1]{Arviandy G. Aribowo \ORCID{0000-0003-3891-3123}}
\author[1]{Ramachandran Anantharaman \ORCID{0000-0003-4205-582X}}
\author[1]{César A. Gómez-Pérez \ORCID{0000-0001-5649-7608}}
\author[1,$\ast$]{Leyla Özkan \ORCID{0000-0001-8442-772X}}

\authormark{C. Portilla et al.}

\address[1]{\orgdiv{Department of Electrical Engineering}, \orgname{Eindhoven University of Technology}, \orgaddress{\street{P.O.Box 513}, \postcode{5600 MB}, \state{Eindhoven}, \country{The Netherlands}}}

\corresp[$\ast$]{Corresponding author. \href{email:l.ozkan@tue.nl}{l.ozkan@tue.nl}}




\abstract{This paper explores the application of data-driven system identification techniques in the frequency domain to obtain simplified, control-oriented models of photosynthesis regulation under oscillating light conditions. In-silico datasets are generated using simulations of the physics-based Basic DREAM Model (BDM) \cite{Fuente2024PlantPhysio}, with light intensity signals---comprising DC (static) and AC (modulated) components as input and chlorophyll fluorescence (ChlF) as output. Using these data, the Best Linear Approximation (BLA) method is employed to estimate second-order linear time-invariant (LTI) transfer function models across different operating conditions defined by DC levels and modulation frequencies of light intensity. Building on these local models, a	Linear Parameter-Varying (LPV) representation is constructed, in which the scheduling parameter is defined by the DC values of the light intensity, providing a compact state-space representation of the system dynamics.}
\keywords{Photosynthesis regulation, Oscillating light, Frequency-domain system identification, Best linear approximation, Linear-parameter varying model.}


\maketitle

\section{Introduction}
\input{Section_1_2_Introduction}

\section{Basic DREAM Model (BDM)}\label{sec:Basic-DREAM-Model}
\input{Section_2_1_BDM.tex}

\section{Frequency-domain system identification: Best Linear Approximation (BLA) method}\label{sec:Frequency-domain-system-ident}
\input{Section_3_1_FrequencyDomainSI.tex}

\section{Linear parameter-varying (LPV) model development}\label{sec:Linear-Parameter-Varying-(LPV)}
\input{Section_4_4_LPV.tex}

\section{Conclusion}\label{sec:Conclusion}
\input{Section_5_1_Conclusion.tex}


\begin{appendices}
	\section{Appendix A: Detailed description of BDM}\label{sec:Detailed-BDM}
	\input{Section_6_1_Appendix01_BDM.tex}

\end{appendices}

\section{Competing interests}
No competing interest is declared.

\section{Author contributions statement}
C.P. is for writing -- original draft, visualization, validation, software, methodology, and conceptualization. 
A.G.A, R.A., and C.A.G.-P. are for writing -- review and editing. 
L.Ö. is for writing -- review and editing, supervision, methodology, and conceptualization.

\section{Acknowledgments}
The work was funded by HORIZON-EIC-2021-PATHFINDER OPEN project DREAM with the grant agreement no. 101046451. The authors would like to acknowledge David Fuente, Dusan Lazar and Ladislav Nedbal for their technical support in relation to the Basic Dream Model.

\bibliographystyle{abbrvnat} 
\bibliography{DREAM_Bib}


%

\end{document}

%% file: Section_1_2_Introduction.tex

Cultivating plants demands vast resources---about
70\% of all water used in Spain, and roughly 7\% of national energy
consumption for greenhouses in the Netherlands. With mounting environmental
and other pressures, current cultivation practices are under strain.
A major challenge for the coming decades is to deploy breakthrough,
high-throughput technologies to improve plants---both with and without
genome engineering---and to enable more resource-efficient cultivation.
This is the vision behind DREAM, an EU-funded, science-driven technology
initiative.\footnote{See more information about the EU-funded DREAM initiative project
in \href{https://dream-eic.eu/}{https://dream-eic.eu/}}

To optimize plant growth, it is desirable to enhance lighting efficiency
\cite{Baker2008AnnualReviewChlorophyllFluorsce}, which can be performed
by dynamically modelling and optimizing the light acclimation of photosynthetic
organisms. The aim is to exploit mathematical models to tailor time-modulated
illuminations such that they can optimize the photon budget (enhanced
lighting efficiency) for driving photosynthesis (i.e., in controlled
environments). However, actual light optimization requires the understanding
and prediction of photon used by PSII. In order to achieve these goals,
the DREAM consortium is undertaking experimental studies with organisms,
such as algae and tomatoes, and also developing dynamic models (based
on the first-principle and experimental data) to study the photosynthesis
regulation under modulated light conditions.

Photosynthesis is a complex process that strongly influences plant
behavior. It involves mechanisms operating on different time scales,
ranging from fast dynamics in light harvesting to slow dynamics in
the chemical synthesis of organic compounds for plant growth. Moreover,
the photosynthetic process exhibits dynamic responses to unfavorable
conditions or to its own physiological state, such as photorespiration
\cite{Ogawa1982SimpleOscillPhoto}, photoinhibition \cite{EilersandPeeters1988};
\cite{WuandMerchuk2001}, and light acclimation processes \cite{DenmanandMarra1986}.
Developing a complete model of photosynthesis would require significant
effort and extensive experimental data. A more practical approach
is to focus on specific characteristic phenomena. The literature contains
numerous models that aim to predict photosynthetic behavior. Some
of these address: $\textrm{CO}_{2}$ assimilation in the Calvin cycle
\cite{Pearcyetal1997}, the activities of the Rubisco enzyme (associated
with the consumption of $\textrm{CO}_{2}$ and $\textrm{O}_{2}$)
and the delayed $\textrm{CO}_{2}$ release during photosynthetic assimilation
and photorespiration processes in leaves \cite{Roussel2011DynMechOscilPhoto},
and light acclimation and photoinhibition under periodically changing
light \cite{PahlandImboden1990}. Light optimization and control in
greenhouses are related to the light-harvesting function of photosystem
II (PSII). The efficiency of light used by plants has been evaluated
through chlorophyll fluorescence \cite{Baker2008AnnualReviewChlorophyllFluorsce},
which provides indirect measurements of photochemical and non-photochemical
quenching. This concept underpins the Basic Dream Model (BDM) developed
for photosynthetic reactions that depend on light intensity oscillations
between $10^{-3}$ and $10^{3}$ Hz, as reported in \cite{Fuente2024PlantPhysio}.
Here, the authors propose a mechanistic representation of electron
transport in PSII and a dynamic model to predict photochemical and
non-photochemical quenching under harmonically oscillating light.
This model is suitable for light optimization; however, it remains
too complex for control design purposes.

Oscillating light is closely related to photochemical and non-photochemical
quenching mechanisms that plants have developed to enhance energy
use. Examples of naturally occurring oscillating light patterns include
leaf flutter and the movement of algal cells in turbulent flows. The
study in \cite{Salvatori2022SysDynApproachPhotosyn} introduced and
validated a simple dynamic model that captures how photosynthesis
responds to fluctuating light at the leaf level, bridging experimental
data and theory. By highlighting differences between genotypes and
emphasizing the role of dynamic processes like non-photochemical quenching
(NPQ) and cyclic electron flow, it provides a foundation for scaling
up to canopy-level models of photosynthesis in realistic environments.
The works \cite{Nedbal2021PhotosynDyn,Lazar2022InsightsRegulationPhotosyn,Niu2023PlantsCope}
pioneered the use of frequency-domain analysis (already well-established
in the field of control systems) as an alternative approach in photosynthesis
studies under oscillating light experiments. They reveal how plants
regulate photosynthetic regulation and electron transport processes
across different temporal scales, providing mechanistic insight into
how photosynthesis is stabilized under natural, fluctuating light. 

\textcolor{black}{Understanding all these phenomena are fundamental for analysis of dynamical
behavior of biochemical processes in plants, which leads to improved performance of associated reactions. Although various models have been proposed to describe the influence of oscillating light on photosynthesis, their inherently nonlinear mathematical structures often are too complex for practical applications of regulation of photosynthesis. For example,
the nonlinear terms in these models can hinder development and use of standard control design tools for light modulation. Alternatively, system identification methods are a data-driven modeling approach that can capture key system dynamics with reduced complexity, offering a more practical route for investigating and optimizing photosynthetic processes.}

System identification techniques, widely used in engineering and control
theory, have not yet been broadly applied to study and analyze photosynthesis
regulation in plants. Their integration could provide powerful tools
to quantitatively model, predict, and control dynamic responses in
photosynthetic systems, but this potential remains largely unexplored.
For example, \cite{Bala2025SysIdChloroFluor} demonstrates that a
data-driven system identification approach can effectively model plant
chlorophyll fluorescence responses under dynamic light, bridging control
engineering and plant physiology. By showing that black-box models
can approximate biologically meaningful processes, it lays the groundwork
for real-time phenotyping, stress detection, and digital twin applications
in agriculture.

Advanced system identification techniques for nonlinear systems, such
as the Best Linear Approximation (BLA) framework and Linear Parameter-Varying
(LPV) models, have been extensively developed and refined in control
engineering to capture nonlinearities, process noise, and varying
operating conditions. However, these powerful data-driven approaches
have not yet been systematically explored for modeling and analyzing
photosynthesis regulation. The BLA method, see further in \cite{Dobrowiecki2007LinearApproxWeaklyNonlinear,Pintelon2013FRFMeasNonlinearSysCL,Schoukens2018IFACDetectingNonlinearModulsDynamicNetwork,Schoukens2019NonlinearSysIdUserRoadmap,Schoukens2020ExtendingBLAProcessNoise},
provides a systematic way to approximate nonlinear systems by a linear
model that best represents their dynamics under specific excitation
conditions, while also quantifying nonlinear distortions and noise
contributions. Its significance lies in enabling robust and interpretable
models of nonlinear dynamics, serving as a bridge toward more complex
nonlinear identification. While the LPV modeling framework (see, for
example, \cite{LaurainToth2010AutomaticaIVMethodLPV,Lovera2013LPVModelIdentificationOverview,BachnasToth2014JPCDataDrivenLPV}),
by contrast, generalizes linear system descriptions by allowing model
parameters to vary with measurable scheduling variables, thus capturing
a wide range of nonlinear and time-varying behaviors. This significantly
enables scalable, data-driven representations of nonlinear systems
that remain compatible with powerful linear control and identification
tools, making it highly suitable for complex biological regulation
problems.

In this paper, we present modeling results demonstrating the application
of a frequency-domain system identification approach (i.e., in terms
of transfer functions) that capture the input--output dynamics of
photosynthetic regulation. The main contributions of this work can
be summarized as follows: 
\begin{enumerate}
\item Application of BLA for local modeling: We employ the BLA method on
in-silico input--output data generated from BDM simulations to construct
local transfer function models under varying operational conditions.
These conditions are defined by the input (light intensity) dynamics,
including DC/constant levels as well as modulation frequencies and
amplitudes. 
\item Development of an LPV representation: Building on the local models,
we develop a Linear Parameter-Varying (LPV) representation in which
the scheduling parameter is defined by the operational range of DC
light intensity values, thereby enabling a compact, global model description
across multiple operating regimes. 
\end{enumerate}
\textcolor{black}{Overall, these contributions provide an advanced
data-driven approach for quantitatively characterizing regulatory
mechanisms in photosynthesis. This paper also provides a concise tutorial
on  the application of system identification technique commonly used
in the control systems community }

The remainder of this paper is structured as follows. Section \ref{sec:Basic-DREAM-Model}
introduces briefly the BDM as the foundation for generating in-silico
input--output data. Section \ref{sec:Frequency-domain-system-ident}
details the application of the BLA method for deriving local transfer
function models under varying light conditions. Section \ref{sec:Linear-Parameter-Varying-(LPV)}
then presents the construction of an LPV representation to capture
system behavior across operating regimes. Finally, Section \ref{sec:Conclusion}
summarizes the key findings and discusses an outlook for future research
on system identification in photosynthesis regulation.

%% file: Section_2_1_BDM.tex
The Basic Dream Model (BDM) represents dynamical systems of photosynthetic
reactions under the influence of light intensity oscillations within
the frequency range of $10^{-3}$ -- $10^{3}$ Hz \cite{Fuente2024PlantPhysio}.
The BDM is described by a set of nonlinear ordinary differential equations
(ODE), which can be formed as the following state and output equations
(i.e., also known as the state-space form): 
\begin{subequations}
\label{eq:nonlinear_bdm_model}

\begin{align}
\dot{\mathbf{x}}(t) & =f(\mathbf{x}(t),u(t),\mathbf{\Phi}),\label{eq:x_dot}\\
\mathbf{y}(t) & =g(\mathbf{x}(t),u(t),\mathbf{\Phi}).\label{eq:y}
\end{align}
\end{subequations}
 where the state variables $\mathbf{x}(t)\in\mathbb{R}^{5}$ with 
\begin{itemize}
\item $x_{1}(t)=\textrm{PQ(\ensuremath{t})}$ for the oxidized plastoquinone
pool, 
\item $x_{2}(t)=H_{L}(t)$ for the proton concentration in the thylakoid
lumen, 
\item $x_{3}(t)=FQ_{act}(t)$ for the active fraction of the photosystem
II quencher, 
\item $x_{4}(t)=\textrm{ATP}(t)$ for the ATP concentration in the stroma, 
\item $x_{5}(t)=\textrm{PI}_{ox}(t)$ for the photosystem I donors that
accept electrons from the plastoquinone pool. 
\end{itemize}
The vector field $f(.)$ in Eq. \eqref{eq:x_dot} describes the evolution
of the state vector $\mathbf{x}(t)$. The vector field $g(.)$ in
Eq. \eqref{eq:y} gives the trajectory of the output vector $\mathbf{y}(t)$;
see further details about these vector fields in Appendix \ref{sec:Detailed-BDM}.
The input signal $u(t)=L(t)$ is the light intensity composed of two
components as follows: the DC/static signal and the modulation signal
(in terms of the amplitude and frequency of light oscillation). The
output vector $\mathbf{y}(t)\in\mathbb{R}^{3}$ is composed by 
\begin{itemize}
\item $y_{1}(t)=\textrm{ChlF}$, for the chlorophyll fluorescence yield, 
\item $y_{2}(t)=\textrm{NPQ}$, for the non-photochemical quenching, 
\item $y_{3}(t)=\frac{\textrm{d}O_{2}}{\textrm{d}t}$, for the rate of oxygen
evolution due to the water splitting in PSII. 
\end{itemize}
This study considers only the chlorophyll fluorescence output $y_{1}(t)$
to generate in-silico data for system identification, as detailed
in the next section. For simplifying the notation, now we write the
chlorophyll fluorescence output $y_{1}(t)$ as $y(t)$.

%% file: Section_3_1_FrequencyDomainSI.tex
\noindent Section \ref{subsec:NP_analysis} details the methodology
for obtaining the Frequency Response Function (FRF) for the BDM using
nonparametric frequency domain analysis. Combined with carefully designed
excitation signals, this approach effectively separates noise from
nonlinearities. Furthermore, Section \ref{subsec:P_analysis} focuses
on identifying a parametric model within the frequency domain based
on the nonparametric analysis.

\subsection{Non-parametric analysis}

\label{subsec:NP_analysis}

\subsubsection{The Frequency Response Function (FRF)}

\noindent The FRF represents the relationship between the input and
the output in the frequency domain and is an experimental (data) representation
of the transfer function $G(j\omega)$. For linear systems, the transfer
function $G(j\omega)$ is given by:

\begin{equation}
G(j\omega)=\frac{Y(j\omega)}{U(j\omega)},\label{eq:Linear_TF}
\end{equation}

\noindent where $Y(j\omega)$ and $U(j\omega)$ are the Fourier transform
(FT) of the output signal $y(t)$ and the input signal $u(t)$. For
a periodic excitation signal $u(t)=A\,\sin(\omega t+\phi)$ with frequency
$\omega$ and phase $\phi$, the output is calculated as: $y(t)=|G(j\omega)|$
$A\,sin(\omega t+(\phi+\angle G(j\omega)))$, where $|G(j\omega)|$
and $\angle G(j\omega)$ are the magnitude and phase of $G(j\omega)$,respectively.
However, photosynthetic organisms exhibit non-linear dynamical behavior,
and the fluorescence measurements are affected by noise. In this case,
Eq.\eqref{eq:Linear_TF} can be rewritten as follows:

\begin{equation}
Y(k)=G(j\omega)\,U(k)+\sigma_{NL}^{2}(k)+\sigma_{n}^{2}(k),
\end{equation}

\noindent where $\sigma_{NL}^{2}(k)$ and $\sigma_{n}^{2}(k)$ represent
the distortions caused by nonlinearities and noise, respectively.

\subsubsection{Multi-frequency (multisine) excitation}

\noindent The frequency response function (FRF) can be obtained using
stepped sine signals, where the system is excited by varying the light
intensity according to a sinusoidal function, $\sin(2\pi ft)$, at
a modulation frequency $f$ at a time instant $t$. However, this
approach is time-consuming because transients must reach steady state
before data can be collected at each subsequent frequency. As a more
efficient alternative, multi-frequency (multisine) excitation signals
can be used, comprising a set of $F$ frequencies and defined as follows:

\begin{equation}
u(t)=u_{dc}+\sum_{k\in F}A_{k}\sin(2\,\pi\,f_{k}\,t+\phi_{k}),\label{eq:multisine}
\end{equation}

\noindent where $A_{k}$, $f_{k}$, and $\phi_{k}$ are the amplitude,
frequency, and phase of the modulation, respectively. The frequencies
in the set $F$ can be uniformly distributed in the frequency range
of interest or follow a logarithmic distribution, depending on the
desired resolution. When the amplitudes $A_{k}$ are set equal for
all frequencies in the set $F$, the signal power is uniformly distributed
throughout the frequency spectrum. A DC offset $u_{dc}$ was added
to ensure nonnegative light intensity.

\begin{figure}[h]
\centering{}\includegraphics[scale=0.8]{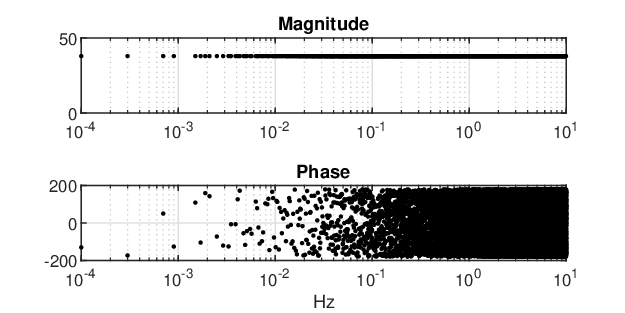}\caption{Spectrum of multisine signal with random phase and amplitude of $38\mu Em^{-2}s^{-1}$
\cite{PintelonSchoukens2012BookSystemIdentification,Schoukens2012BookMasteringSysId}.}
\label{fig:multisine_spectrum} 
\end{figure}

\noindent In the simulation, $u_{dc}$ was tuned to enhance signal
power with minimal oscillations, for example, using the modulation
amplitude $A_{k}=38\mu Em^{-2}s^{-1}$ and $F=30000$ frequencies
uniformly distributed from $10^{-4}$ to $10$ Hz as in the magnitude
plot Figure \ref{fig:multisine_spectrum}. Note that the modulation
amplitudes in this magnitude plot are identical in all frequencies.
Furthermore, the phases $\phi_{k}$ are randomly assigned to avoid
large amplitude peaks in the multisine signal, thereby reducing the
crest factor and improving the noise-to-signal ratio (NSR). A detailed
discussion of crest factor analysis is provided in \cite{PintelonSchoukens2012BookSystemIdentification,Schoukens2012BookMasteringSysId}.

\subsubsection{Noise and nonlinear distortions characterization}

Using non-parametric frequency domain analysis, we obtained the Frequency
Response Function (FRF) through carefully designed excitation signals,
which separated noise from nonlinearities. Discriminating between
these requires careful experimental planning, especially by selecting
excitation signals. Periodic signals are particularly useful during
non-parametric preprocessing (without considering interactions), as
they allow the calculation of the FRF, noise power spectrum, and nonlinear
distortions \cite{Schoukens2019NonlinearSysIdUserRoadmap}. Note that
we focus on nonlinear PISPO (Period-In, Same-Period-Out) systems,
where the steady-state response to a periodic input is expected to
produce an output signal within the same period.

\noindent To characterize the noise, the system should be excited
for at least two periods after reaching a steady state. This method
is effective because, while noise varies between periods, nonlinear
distortions remain consistent \cite{Schoukens2019NonlinearSysIdUserRoadmap}.

Let us consider a system $y(t)=f(u(t))$ excited with a periodic signal
$u(t)$ during $P$ periods in steady-state conditions. The Discrete
Fourier Transform (DFT) is calculated for the input and output of
each period, obtaining $U^{[l]}(k)$, $Y^{[l]}(k)$, for $l=1,...,P$.
The sample mean for each experiment $\hat{Y}(k)$ and noise variance
$\sigma_{Y}^{2}(k)$ are given by:

\begin{align}
\hat{Y}(k) & =\frac{1}{P}\sum_{l=1}^{P}Y^{[l]}(k),\\
\sigma_{Y}^{2}(k) & =\frac{1}{P-1}\sum_{l=1}^{P}|Y^{[l]}(k)-\hat{Y}(k)|.
\end{align}

\noindent Nonlinearities can be assessed by repeating the experiment
using different realizations of random-phase multisine excitation.
To quantify nonlinearities, the principle involves calculating the
variance, $\sigma_{NL}^{2}(k)$. In linear systems, the phase relationship
between input and output remains constant. Using this property and
generating multiple realizations of random phase multisine signals,
the variance can be used to estimate non-linear distortions. In general,
for the FRF for the experiment $m$ is given by:

\begin{equation}
G^{[m]}(j\omega_{k})=\frac{\hat{Y}^{[m]}(k)}{\hat{U}^{[m]}(k)}.
\end{equation}

\noindent For a set of $M$ experiments using $M$ realizations of
a random phase multisine signal, the Best Linear Approximation (BLA)
is given by \cite{PintelonSchoukens2012BookSystemIdentification,Schoukens2012BookMasteringSysId}

\begin{equation}
G_{BLA}(j\omega_{k})=\frac{1}{M}\sum_{m=1}^{M}(j\omega_{k}).
\end{equation}

\noindent The total distortion, accounting for both nonlinearities
and noise, can then be calculated as follows:

\begin{equation}
\sigma_{G_{BLA}}^{2}(j\omega_{k})=\sum_{m=1}^{M}\frac{|G^{[m]}(j\omega_{k})-G_{BLA}(j\omega_{k})|^{2}}{M(M-1)}.
\end{equation}

\begin{figure}[h]
\centering{}\includegraphics[scale=0.8]{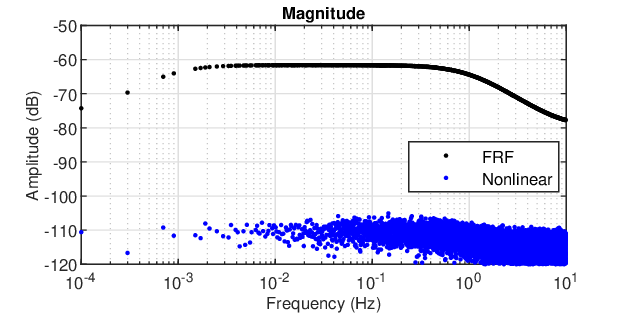} \caption{Magnitude spectrum of fluorescence in response to multisine light
intensity with an amplitude of $38\mu E,m^{-2},s^{-1}$ and a DC value
of $100\mu E,m^{-2},s^{-1}$. The plot distinguishes between linear
and nonlinear behavior.}
\label{fig:nonparametric} 
\end{figure}

\noindent Applying the multisine light intensity shown in Figure \ref{fig:multisine_spectrum}
and adding a DC value of $100\mu E,m^{-2},s^{-1}$, \textcolor{black}{the variance between
the last four periods (during the steady state) and the six realizations
of the multisine signal can be calculated.} This discriminates FRF
and non-linear behavior, as illustrated in Figure \ref{fig:nonparametric}.
Note that this example has no noise, as each period is identical to
the others. This is because the data are derived from a simulation
of the BDM.

\noindent Figure \ref{fig:nonparametric} shows that the separation
between the Frequency Response Function (FRF) and the nonlinearities
exceeds $40$ dB, indicating that the linear contribution is 100 times
stronger than the nonlinearities across all frequencies.

\subsection{Parametric analysis}

\label{subsec:P_analysis}

\noindent For performing parametric identification in the frequency
domain (see further in \cite{PintelonSchoukens2012BookSystemIdentification,Schoukens2012BookMasteringSysId}),
consider that a transfer function $G(z_{k})$ at frequency $k2\pi/N$
is given by:

\begin{equation}
G(z_{k})=\frac{Y(k)}{U(k)}=\frac{B(z_{k})}{A(z_{k})}=\frac{\sum_{l=0}^{n_{b}}b_{k}z_{k}^{-l}}{\sum_{l=0}^{n_{a}}a_{k}z_{k}^{-l}},
\end{equation}

\noindent with $z_{k}$ being the discrete domain frequency variable
at frequency $kf_{0}$. The objective is to find the polynomials coefficients
of $B(z_{k})$ and $A(z_{k})$ (i.e., the values of $b_{k}$ and $a_{k}$)
and approximate the polynomial degrees $n_{b}$ and $n_{a}$.

\begin{align}
a_{0}Y(k)+a_{1}Y(k)z_{k}^{-1}+...+a_{(}n_{a})Y(k)z_{k}^{-n_{a}}=\cdots\nonumber \\
\cdots b_{0}U(k)+b_{1}U(k)z_{k}^{-1}+...+b_{(}n_{b})Y(k)z_{k}^{-n_{b}}.
\end{align}
Assuming $a_{0}=1$, a static set of complex equations can be constructed
as:

\begin{equation}
\underbrace{\begin{bmatrix}Y(0)\\
\vdots\\
Y(N/2)
\end{bmatrix}}_{\mathbf{Y}}=\underbrace{\begin{bmatrix}Y(0)z_{0}^{-1} & \hdots & Y(N/2)z_{N/2}^{-1}\\
\vdots & \ddots & \vdots\\
Y(0)z_{0}^{n_{a}} & \hdots & Y(N/2)z_{N/2}^{-n_{a}}\\
U(0) & \hdots & U(N/2)\\
\vdots & \ddots & \vdots\\
U(0)z_{0}^{-nb} & \hdots & U(N/2)z_{N/2}^{-n_{b}}
\end{bmatrix}^{\mathsf{T}}}_{\mathbf{K}}\underbrace{\begin{bmatrix}a_{1}\\
\vdots\\
a_{n_{a}}\\
b_{0}\\
\vdots\\
b_{n_{b}}
\end{bmatrix}}_{\mathbf{\theta}}.
\end{equation}

\noindent The linear system above is given by $Y=K\theta$, where
$\theta$ represents the vector of unknown parameters. The previous
complex system can be transformed into a real-valued system by:

\begin{equation}
\begin{bmatrix}\mathrm{real}(\mathbf{Y})\\
\mathrm{imag}(\mathbf{Y})
\end{bmatrix}=\begin{bmatrix}\mathrm{real}(\mathbf{K})\\
\mathrm{imag}(\mathbf{K})
\end{bmatrix}\mathbf{\theta}.
\end{equation}

\noindent Then, the previous system $\mathbf{Y_{real}}=\mathbf{K_{real}}\,\mathbf{\theta}$
can be solved using Least Squares (LS) as:

\begin{equation}
\mathbf{\theta}=(\mathbf{K_{real}}{\T}\mathbf{K_{real}})^{-1}\mathbf{K_{real}}{\T}\mathbf{Y_{real}}.\label{eq:LS}
\end{equation}

\begin{figure}[h]
\centering{}\includegraphics[scale=0.8]{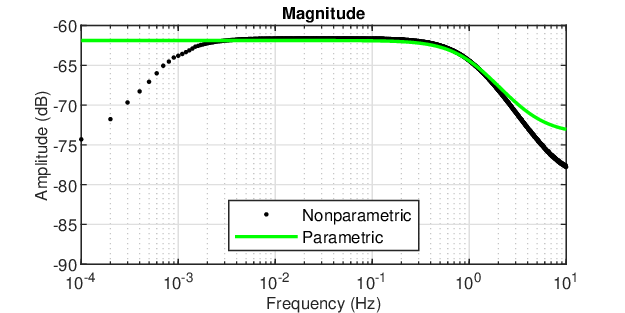}
\caption{Comparison of nonparametric and parametric (LS) models for fluorescence
response to multisine light intensity (amplitude: $38\mu Em^{-2}s^{-1}$,
DC value: $100\mu Em^{-2}s^{-1}$).}
\label{fig:LS_parametric} 
\end{figure}

\noindent At this stage, the user must decide on the complexity of
the linear model by selecting the degrees of $B(z_{k})$ and $A(z_{k})$,
that is, determining precisely the values of $n_{b}$ and $n_{a}$.
Generally, a good model adheres to the principle of parsimony, achieving
the highest possible accuracy with the fewest parameters. In this
regard, we choose the following polynomial degrees: $n_{b}$=$n_{a}$=2.
As a result, a second order model is derived from the nonparametric
analysis presented in Figure \ref{fig:nonparametric}. The magnitude
spectra of the parametric (also called the magnitude bode plot) and
nonparametric models are shown in Figure \ref{fig:LS_parametric}.
The Least Squares approximation fails for frequencies below $10^{-3}$
Hz due to the scarcity of data points at these frequencies. Consequently,
the solution neglects this frequency range.

\noindent A weighted version of the Least Squares (WLS) method is
implemented to address the scarcity of data issue mentioned above,
with an increased emphasis on lower frequencies. Consequently, Eq.
\eqref{eq:LS} can be rewritten as follows:

\begin{equation}
\mathbf{\theta}=(\mathbf{K_{real}}{\T}\,\mathbf{W}\,\mathbf{K_{real}})^{-1}\mathbf{K_{real}}{\T}\,\mathbf{W}\,\mathbf{Y_{real}}.\label{eq:WLS}
\end{equation}

\begin{figure}[h]
\centering{}\includegraphics[scale=0.8]{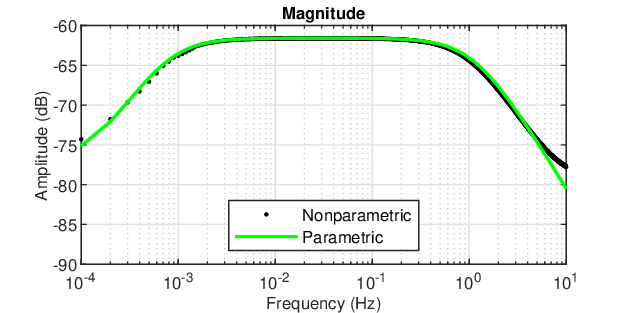}
\caption{Comparison of nonparametric and parametric (WLS) models for fluorescence
response to multisine light intensity (amplitude: $38\mu Em^{-2}s^{-1}$,
DC value: $100\mu Em^{-2}s^{-1}$).}
\label{fig:WLS_parametric} 
\end{figure}

\noindent Figure \ref{fig:WLS_parametric} shows the results of the
Weighted Least Squares (WLS) method, where the parametric model achieves
a good fit across both low and high frequencies.

\noindent The next section explains how a set of linear models is
used to represent nonlinear behavior through a Linear Parameter-Varying
(LPV) modeling approach.

%% file: Section_4_4_LPV.tex
\noindent An LPV model representation for the in-silico data is developed
based on a local approach. First, we present a set of local linear
parametric models based on the FRF of the nonlinear system in Section
\ref{subsec:LPV-model-structure}. Section \ref{subsec:Parametrization-of-LPV}
formalizes the proposed scheduling variable for the LPV model. Section
\ref{subsec:Space-state-representation-LPV} outlines the state space
representation of the proposed LPV model. Section \ref{subsec:Simulation-results-LPV}
validates this LPV model via numerical simulations.

\subsection{LPV model structure} \label{subsec:LPV-model-structure}

\begin{figure}[h]
\centering{}\includegraphics[scale=0.3]{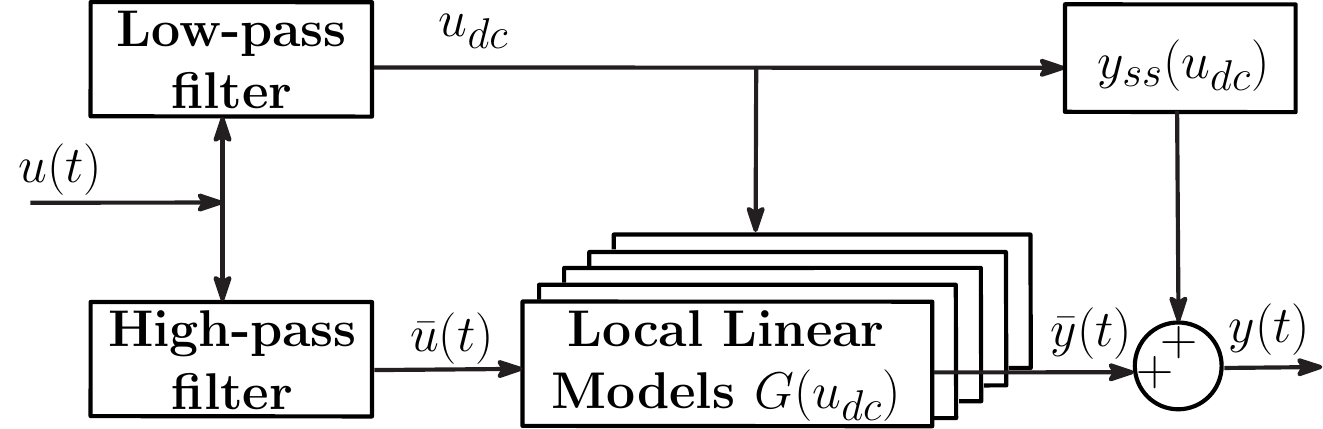}
\caption{LPV model structure. }\label{fig:LPVstructure}
 
\end{figure}

By adopting a local modeling approach, the construction of an LPV
model is based on a set of linear models that is interpolated based
on a scheduling variable, as illustrated in Figure \ref{fig:LPVstructure}.
In this framework, the parameters of the local linear models are parameterized
as functions of the scheduling variable. Specifically in our case,
the scheduling variable is the DC component of the light intensity
signal ($u(t)$), which can either be determined offline or estimated
in real-time using a low-pass filter.

\noindent Note that, for a linear model, its input and output are
expressed in terms of perturbation coordinates. Hence, the perturbation
input variable $\bar{u}(t)$ can be calculated by subtracting the
original input variable $u(t)$ from its mean value (i.e., corresponding
to the DC component of the light intensity $u_{dc}$), or alternatively,
it can be obtained in real time using a high-pass filter. For the
output (chlorophyll fluorescence) of BDM (see in Section \ref{sec:Basic-DREAM-Model}),
the measured (original) output signal is given by

\begin{equation}
y(t)=\bar{y}(t)+y_{ss}(u_{dc}),\label{eq:output_y}
\end{equation}

\noindent where the DC output $y_{ss}(u_{dc})$ is parameterized in
terms of a scheduling variable in the LPV model structure. In this
work, the DC light intensity $u_{dc}$ is set as the scheduling variable.

\subsection{Local linear models} \label{subsec:LPV-local-linear-models}

\begin{figure}[t!]
	\centering
	\subfloat[$u_{dc} = 200\,\mu\mathrm{Em}^{-2}\mathrm{s}^{-1}$]{\includegraphics[width=.22\textwidth]{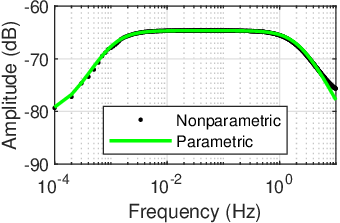}} 
	\hfill
	\subfloat[$u_{dc} = 400\,\mu \mathrm{Em}^{-2}\mathrm{s}^{-1}$]{\includegraphics[width=.22\textwidth]{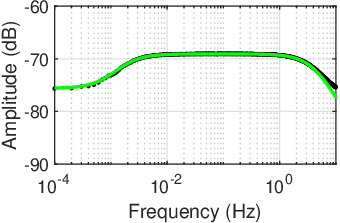}}
	
	\vskip \baselineskip
	\subfloat[$u_{dc} = 600\,\mu \mathrm{Em}^{-2}\mathrm{s}^{-1}$]{\includegraphics[width=.22\textwidth]{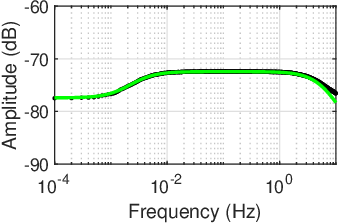}}
	\hfill
	\subfloat[$u_{dc} = 1000\,\mu \mathrm{Em}^{-2}\mathrm{s}^{-1}$]{\includegraphics[width=.22\textwidth]{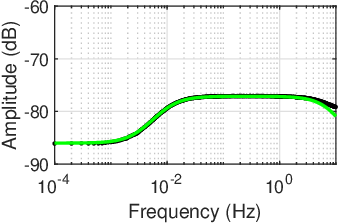}}
	
	\caption{Comparison of the FRF and the magnitude bode plots for parametric models for $u_{dc} = 200, 400, 600$ and $1000\,\mu \mathrm{Em}^{-2}\mathrm{s}^{-1}$.} \label{fig:bode_results2}
\end{figure}

\noindent Following the non-parametric analysis outlined in the previous
section, the Frequency Response Function (FRF) is calculated for various
DC values ($u_{dc}$). The DC values were varied from $100$ to $1000$
in increments of $50$, generating 18 second-order linear and time-invariant
models. Figure \ref{fig:bode_results2} presents the resulting magnitude
bode plots for $u_{dc}=200,\,400,\,600$ and $1000\,\mu E\text{m}^{-2}\text{s}^{-1}$.

\noindent Figure \ref{fig:bode_results2} shows that the DC value
substantially affects the FRF, as evident in the Bode magnitude plot,
where the gain of the system varies with both frequency and changes
in the DC value, affecting the parametric linear model.

\subsection{Parametrization of the scheduling variable $(u_{dc})$} \label{subsec:Parametrization-of-LPV}

\noindent The set of second-order transfer functions computed in previous
sections can be written in continuous-time zero-pole-gain form as:

\begin{equation}
	G(s)=\frac{\boldsymbol{K}\left(s-\boldsymbol{Z_{1}}\right)(s-\boldsymbol{Z_{2}})}{(s-\boldsymbol{P_{1}})(s-\boldsymbol{P_{2}})},\label{eq:TF_zpk}
\end{equation}

\noindent where $\boldsymbol{K}$ is the gain of the system, ${\boldsymbol{Z_{1}},\boldsymbol{Z_{2}}}$
the zeros, and ${\boldsymbol{P_{1}},\boldsymbol{P_{2}}}$ the poles.
The transfer function has equal order in both the numerator and the
denominator, making it a proper transfer function. We guarantee also
that the poles have negative real parts, which ensures stability.

\begin{figure}[h!] 
	\centering{}
	\includegraphics[scale=0.8]{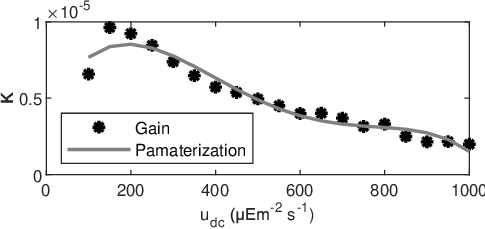} \caption{Parametrization of the system gain with respect to the DC value. }\label{fig:K}
\end{figure}

\noindent The DC value will be used as the scheduling variable for
the LPV model. To achieve this, a parameterization of the DC value
is proposed where we derive functions that describe how the gain ($K(u_{dc})$),
poles ($P_{1}(u_{dc}),P_{2}(u_{dc})$), and zeros ($Z_{1}(u_{dc}),Z_{2}(u_{dc})$)
change as a function of the DC value. The parameterization results
for the gain are presented in Figure \ref{fig:K}, where the coefficients
of a fourth-degree interpolation polynomial were estimated, achieving
an $R^{2}=0.94$.

\begin{figure}[h!] 
	\centering{}
	\includegraphics[scale=0.68]{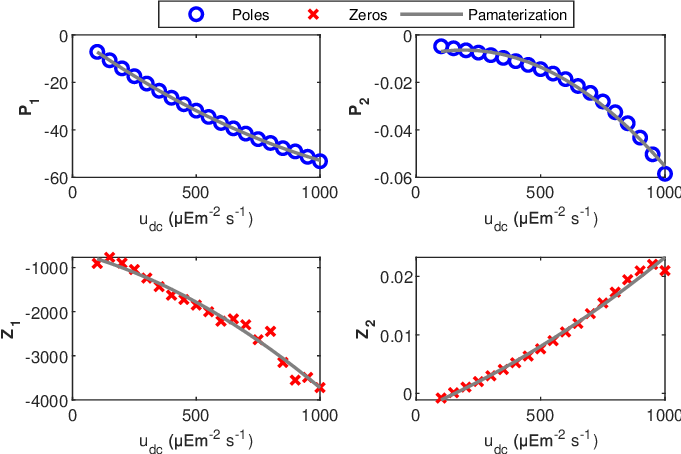} \caption{Parametrization of the poles and zeros as a function of the DC value.}\label{fig:PZ}
\end{figure}

\noindent Likewise, the parameterization results for the poles and
zeros are presented in Figure \ref{fig:PZ}, where the coefficients
of a second-degree interpolation polynomial have been estimated. For
all cases, the fitting exceeded $R^{2}=0.97$.

\begin{figure}[h!] 
	\centering{}
	\includegraphics[scale=0.8]{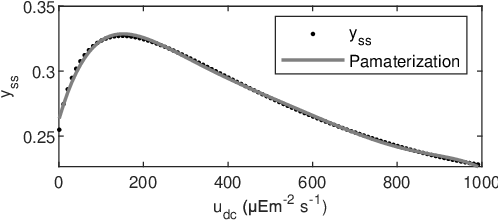} \caption{Parametrization of the steady-state output as a function of the DC value.}\label{fig:yss} 
\end{figure}

\noindent Additionally, it is necessary to parameterize the steady
state value as a function of the scheduling variable $u_{dc}$. To
achieve this, the steady-state response for constant light excitation
was obtained across the same range of DC values. Consequently, the
parameterization was performed using Fourier series up to the third
harmonic. The result is shown in Figure \ref{fig:yss}, where the
fundamental frequency is $f_{0_{ss}}=\cne{6.7}{-5}$ Hz, indicating
that $y_{s}s(u_{dc})$ is being parameterized as a periodic signal.

\noindent Finally, the parameterization can be formulated by following non-linear static functions:

\begin{align}
K(u_{dc}) & =\cne{3.97}{-6}+\frac{u_{dc}}{\cne{5.62}{8}}\ldots\nonumber \\
 & \ldots-\frac{u_{dc}^{2}}{\cne{2.17}{10}}+\frac{u_{dc}^{3}}{\cne{2.76}{13}}-\frac{u_{dc}^{4}}{\cne{1.17}{16}},\nonumber \\
P_{1}(u_{dc}) & =-\cne{1.11}{-2}-\frac{u_{dc}}{\cne{7.48}{2}}+\frac{u_{dc}^{2}}{\cne{2.19}{5}},\nonumber \\
P_{2}(u_{dc}) & =-\cne{9.08}{-3}+\frac{u_{dc}}{\cne{2.82}{5}}-\frac{u_{dc}^{2}}{\cne{7.45}{8}},\nonumber \\
Z_{1}(u_{dc}) & =-637.87-1.49\,u_{dc}-\frac{u_{dc}^{2}}{\cne{1.58}{3}},\nonumber \\
Z_{2}(u_{dc}) & =-\cne{2.98}{-3}+\frac{u_{dc}}{\cne{1.78}{5}}+\frac{u_{dc}^{2}}{\cne{8.46}{9}},\nonumber \\
y_{ss}(u_{dc}) & =a_{0}+a_{1}\,\text{sin}(2\pi f_{0_{ss}}u_{dc})\ldots\nonumber \\
 & \ldots+b_{1}\,\text{cos}(2\pi f_{0_{ss}}u_{dc})+a_{2}\,\text{sin}(4\pi f_{0_{ss}}u_{dc})\ldots\nonumber \\
 & \ldots+b_{2}\,\text{cos}(4\pi f_{0_{ss}}u_{dc})\,+a_{3}\,\text{sin}(6\pi f_{0_{ss}}u_{dc})\ldots\\
 & +b_{3}\,\text{cos}(6\pi f_{0_{ss}}u_{dc}),
\end{align}

\noindent where $a_{0}=-20038,\,a_{1}=7236.8,\,b_{1}=29213,\,a_{2}=-5640.9,$
$b_{2}=-10688,\,a_{3}=1349.2,\,b_{3}=1512.8$.

\subsection{Space state representation of the LPV}\label{subsec:Space-state-representation-LPV}

\noindent The second-order transfer function with parameters varying depending on the DC value of the excitation signal can be formulated as follows:

\begin{equation}
G(s)=\frac{Y(s)}{U(s)}=\frac{b_{0}(u_{dc})\ s^{2}+b_{1}(u_{dc})\ s+b_{2}(u_{dc})}{s^{2}+a_{1}(u_{dc})\ s+a_{2}(u_{dc})},\label{eq:TF_lpv}
\end{equation}

\noindent where,

\begin{align}
b_{o}(u_{dc}) & =K(u_{dc}),\nonumber \\
b_{1}(u_{dc}) & =(-Z_{1}(u_{dc})-Z_{2}(u_{dc}))\,K(u_{dc}),\nonumber \\
b_{2}(u_{dc}) & =Z_{1}(u_{dc})\,Z_{2}(u_{dc})\,K(u_{dc}),\nonumber \\
a_{1}(u_{dc}) & =-P_{1}(u_{dc})-P_{2}(u_{dc}),\nonumber \\
a_{2}(u_{dc}) & =P_{1}(u_{dc})\,P_{2}(u_{dc}).
\end{align}

\begin{figure}[h]
\centering{}\includegraphics[scale=0.3]{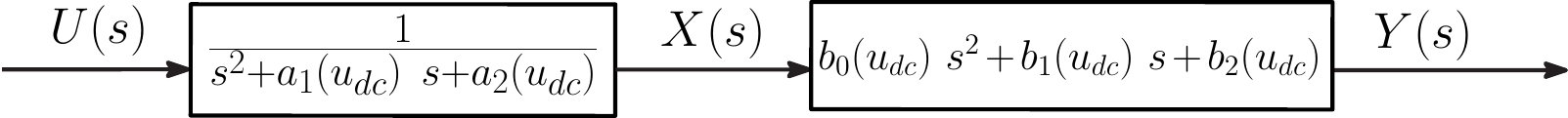} \caption{Transfer function in a block diagram.}
\label{fig:TF2SS} 
\end{figure}

\noindent Eq. \eqref{eq:TF_lpv} can be expressed in a block diagram
as shown in Figure \ref{fig:TF2SS}, where $X(s)$ is called the states
of the system.

\noindent Therefore, the transfer function can be rewritten as:

\begin{equation}
G(s)=\frac{Y(s)}{X(s)}\frac{X(s)}{U(s)},\label{eq:TF2}
\end{equation}

\noindent where $\frac{X(s)}{U(s)}$ shows the dynamic relation between
the input $\bar{u}(t)$ and the state $\bar{x}(t)$ though the differential
equation in time domain:

\begin{equation}
\ddot{\bar{x}}(t)+a_{1}(u_{dc})\dot{\bar{x}}(t)+a_{2}(u_{dc})\bar{x}(t)=\bar{u}(t).\label{eq:eq_x_u}
\end{equation}

\noindent Likewise, $\frac{X(s)}{U(s)}$ in Eq. \eqref{eq:TF2} shows
relation between the state $\bar{x}(t)$ and the output $\bar{y}(t)$
by means of:

\begin{equation}
b_{0}(u_{dc})\ddot{\bar{x}}(t)+b_{1}(u_{dc})\dot{\bar{x}}(t)+b_{2}(u_{dc})\bar{x}(t)=\bar{y}(t).\label{eq:eq_y_x}
\end{equation}

\noindent Assume that $\bar{x}_{1}(t)=\bar{x}(t)$ and $\bar{x}_{2}(t)=\dot{\bar{x}}(t)$.
Then, Eqs. \eqref{eq:eq_x_u}--\eqref{eq:eq_y_x} can be written as a matrix system as:

\begin{align}
\underbrace{\begin{bmatrix}\dot{\bar{x}}_{1}(t)\\
\dot{\bar{x}}_{2}(t)
\end{bmatrix}}_{\mathbf{\dot{\bar{x}}(t)}} & =\underbrace{\begin{bmatrix}0 & 1\\
-a_{2}(u_{dc}) & -a_{1}(u_{dc})
\end{bmatrix}}_{\mathbf{A(u_{dc})}}\underbrace{\begin{bmatrix}{\bar{x}_{1}}(t)\\
{\bar{x}_{2}}(t)
\end{bmatrix}}_{\mathbf{{\bar{x}}(t)}}+\underbrace{\begin{bmatrix}0\\
1
\end{bmatrix}}_{\mathbf{B(u_{dc})}}\bar{u}(t),\nonumber \\
\bar{y}(t) & =\underbrace{\begin{bmatrix}b_{2}(u_{dc})-(b_{0}(u_{dc})a_{2}(u_{dc}))\\
b_{1}(u_{dc})-(b_{0}(u_{dc})a_{1}(u_{dc}))
\end{bmatrix}^{\T}}_{\mathbf{C(u_{dc})}}\underbrace{\begin{bmatrix}{\bar{x}_{1}}(t)\\
{\bar{x}_{2}}(t)
\end{bmatrix}}_{\mathbf{{\bar{x}}(t)}}\ldots\nonumber \\
 & \ldots+\underbrace{\begin{bmatrix}b_{0}(u_{dc})\end{bmatrix}}_{\mathbf{D(u_{dc})}}\bar{u}(t).
\end{align}

\noindent Finally, the state space representation of the linear varying
parameter model is given by:

\begin{align}
\mathbf{\dot{\bar{x}}(t)} & =A(u_{dc})\,\mathbf{\bar{x}(t)}+B(u_{dc})\,\bar{u}(t),\nonumber \\
\bar{y}(t) & =C(u_{dc})\,\mathbf{\bar{x}(t)}+D(u_{dc})\,\bar{u}(t),
\end{align}

\noindent where the state vector $\mathbf{\bar{x}(t)}$ does not have
a physical meaning and $\bar{y}(t)$ is the perturbation variable
of the output (chlorophyll fluorescence) of BDM corresponding to Eq.
\eqref{eq:output_y}.

\subsection{Simulation results}

\label{subsec:Simulation-results-LPV}

\begin{figure}[t!] 
	\centering 
	\subfloat[$f=\cne{1}{-3}$ Hz , $R^{2}=0.88$.\label{fig:DC420_y_equals_x}]{\includegraphics[width=.23\textwidth]{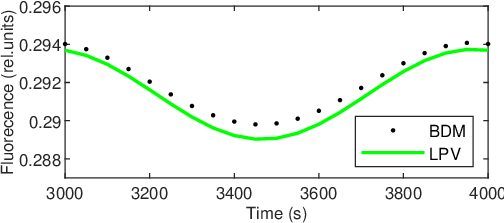}}
	\hfill{}
	\subfloat[$f=\cne{1}{-2}$ Hz, $R^{2}=0.96$.\label{fig:DC420_three_sin_x}]{\includegraphics[width=.23\textwidth]{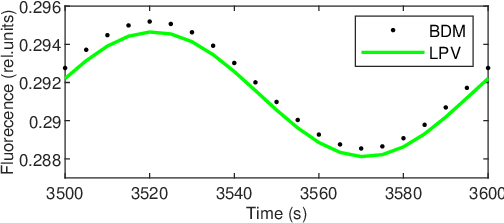}}
	
	\vskip \baselineskip
	\subfloat[$f=\cne{1}{-1}$ Hz, $R^{2}=0.96$.\label{fig:DC420_five_over_x}]{\includegraphics[width=.23\textwidth]{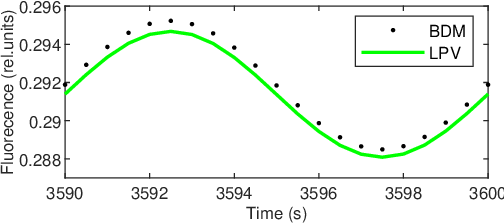}} 
	\hfill{}
	\subfloat[$f=1$ Hz, $R^{2}=0.95$.\label{fig:DC420_five_over_x-1}]{\includegraphics[width=.23\textwidth]{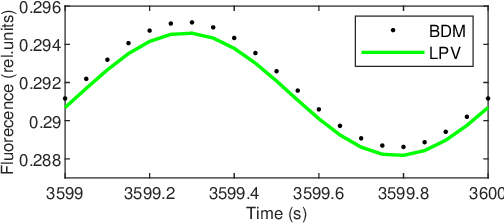}}
	
	\caption{Comparison of fluorescence data at different frequencies, obtained using the BDM and LPV models, for a light oscillating excitation with a DC value of $u_{dc}=420\mu E\text{m}^{-2}\text{s}^{-1}$ and an
		amplitude of $A=10\mu E\text{m}^{-2}\text{s}^{-1}$.}\label{fig:LPVresutls1}
\end{figure}

\noindent In this section, we validate the LPV model through simulations
and comparisons with the nonlinear BDM. Three distinct test cases
are proposed to evaluate the performance of the LPV model under conditions
different from those used during training. Since the LPV model was
developed for steady-state conditions, the data from both the BDM
and the LPV model were collected during the last period of $1$ hour
to ensure that transients had dissipated. The first set of results
is presented in Figure \ref{fig:LPVresutls1}, where the LPV model
is tested at four different frequencies, using an oscillating light
intensity with a DC value of $u_{dc}=420\mu E\text{m}^{-2}\text{s}^{-1}$
and an amplitude of $A=10\mu E\text{m}^{-2}\text{s}^{-1}$.

\begin{figure}[t!]
	\centering
	\subfloat[$f=\cne{1}{-3}$ Hz , $R^{2}=0.88$.\label{fig:DC860_y_equals_x}]{\includegraphics[width=.23\textwidth]{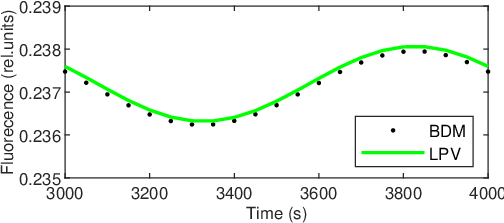}}
	\hfill{}
	\subfloat[$f=\cne{1}{-2}$ Hz, $R^{2}=0.96$.\label{fig:DC860_three_sin_x}]{\includegraphics[width=.23\textwidth]{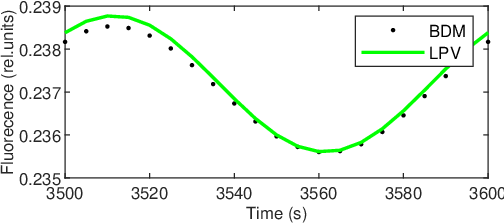}}
	
	\vskip \baselineskip 
	\subfloat[$f=\cne{1}{-1}$ Hz, $R^{2}=0.96$.\label{fig:DC860_five_over_x}]{\includegraphics[width=.23\textwidth]{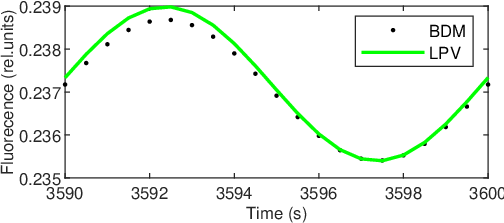}}
	\hfill{}
	\subfloat[$f=1$ Hz, $R^{2}=0.95$.\label{fig:DC860_five_over_x-1}]{\includegraphics[width=.23\textwidth]{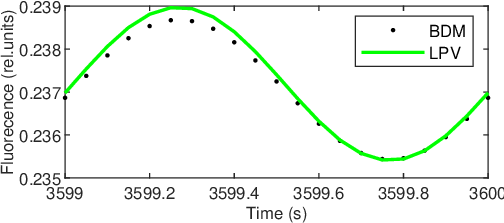}}
		
	\caption{ Comparison of fluorescence data at different frequencies, obtained using the BDM and LPV models, for a light oscillating excitation with a DC value of $u_{dc}=860\mu E\text{m}^{-2}\text{s}^{-1}$ and an amplitude of $A=10\mu E\text{m}^{-2}\text{s}^{-1}$.}\label{fig:LPVresutls2} 
\end{figure}

\noindent Note that the DC value used in this case was not included
in the training data. The results in Figure \ref{fig:LPVresutls1}
show that the LPV model closely replicates the behavior of chlorophyll
fluorescence generated by the BDM, with a $R^{2}$ between $0.88$
and $0.96$. Furthermore, for the lower frequency, $f=\cne{1}{-3}$,
a phase shift occurs, which is accurately captured by the LPV model.
The LPV model was also validated at the same frequencies in a second
case, using a higher DC value ($u_{dc}=860\mu E\text{m}^{-2}\text{s}^{-1}$).
Figure \ref{fig:LPVresutls2} compares the BDM and the LPV model,
demonstrating that the LPV model closely fits the data for non-trained
scenarios. Furthermore, note that the value ranges in Figures \ref{fig:LPVresutls1}
and \ref{fig:LPVresutls2} differ, indicating that the LPV model effectively
tracks changes in magnitude.

\noindent Finally, in the last validation case, the LPV model is tested by applying a multi-frequency light intensity excitation by expanding \eqref{eq:multisine} as follows:

\begin{equation}
	u(t)=u_{dc}+u_{1}(t)+u_{2}(t)+u_{3}(t)+u_{4}(t),
\end{equation}

\noindent where,

\begin{align}
u_{dc} & =530\mu E\text{m}^{-2}\text{s}^{-1},\label{eq:eq_u_t}\\
u_{1}(t) & =20\,\text{sin}(2\pi(\cne{1}{1})t),\nonumber \\
u_{2}(t) & =20\,\text{sin}(2\pi(\cne{1}{-1})t),\nonumber \\
u_{3}(t) & =20\,\text{sin}(2\pi(\cne{1}{-2})t),\nonumber \\
u_{4}(t) & =20\,\text{sin}(2\pi(\cne{1}{-3})t).
\end{align}

\begin{figure}[t!]
	\centering
	\subfloat[Last period of $u(t)$.\label{fig:multisinelight0}]{\includegraphics[width=.23\textwidth]{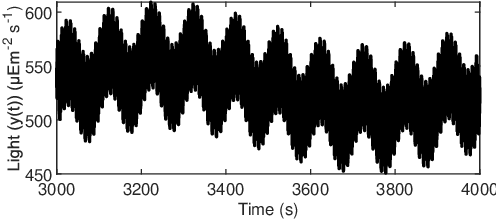}}
	\hfill{}
	\subfloat[Last $100$ seconds of $u(t)$.\label{fig:multisinelight1}]{\includegraphics[width=.23\textwidth]{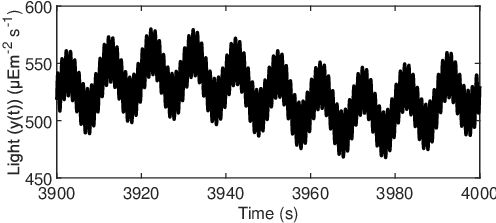}} 
	
	\vskip \baselineskip 
	\subfloat[Last $10$ seconds of $u(t)$.\label{fig:multisinelight2}]{\includegraphics[width=.23\textwidth]{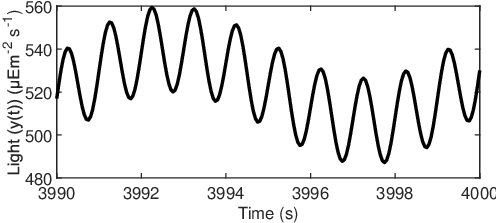}}
	\hfill{}
	\subfloat[Last $1$ second of $u(t)$.\label{fig:multisinelight3}]{\includegraphics[width=.23\textwidth]{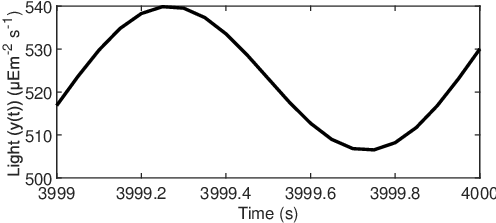}}
	
	\caption{ Comparison of fluorescence data at different frequencies, obtained using the BDM and LPV models, for a light oscillating excitation with a DC value of $u_{dc}=860\mu E\text{m}^{-2}\text{s}^{-1}$ and an
		amplitude of $A=10\mu E\text{m}^{-2}\text{s}^{-1}$.}\label{fig:multisinelight} 
\end{figure}

\noindent Note that the period of the signal $u(t)$ corresponds to the lowest frequency \cn{1}{-3} Hz in Eq. \eqref{eq:eq_u_t}, resulting a period of $1000$ seconds. 
The systems were simulated for a duration of $4000$ seconds to obtain the results, and we collected the chlorophyll fluorescence data from the last period for both the BDM and LPV models.
The light intensity profile $u(t)$ is shown in Figure \ref{fig:multisinelight}, with its complete final period in Figure \ref{fig:multisinelight0}.
Figures \ref{fig:multisinelight1}, \ref{fig:multisinelight2}, and \ref{fig:multisinelight3} present the last $100$, $10$, and $1$ seconds of $u(t)$, respectively, highlighting its direct relationship with the frequencies in $u_{3}$, $u_{2}$, and $u_{1}$.

\begin{figure}[t!]
	\centering 
	\subfloat[Last period of $y(t)$.\label{fig:mulisineresutls1}]{\includegraphics[width=.23\textwidth]{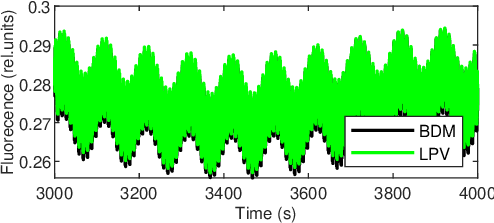}}
	\hfill{}
	\subfloat[Last $100$ seconds of $y(t)$.\label{fig:mulisineresutls2}]{\includegraphics[width=.23\textwidth]{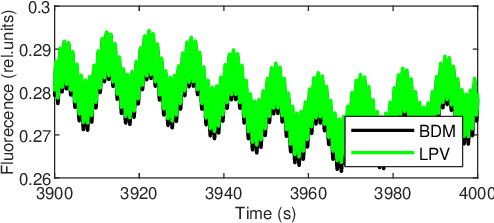}}
	
	\vskip \baselineskip 
	\subfloat[Last $10$ seconds of $y(t)$.\label{fig:mulisineresutls3}]{\includegraphics[width=.23\textwidth]{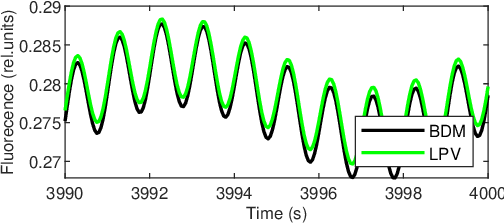}}
	\hfill{}
	\subfloat[Last $1$ second of $y(t)$.\label{fig:mulisineresutls4}]{\includegraphics[width=.23\textwidth]{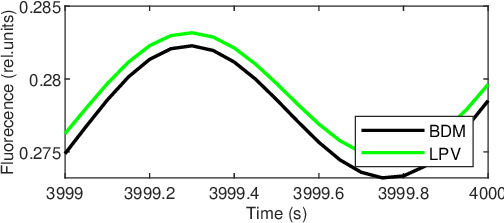}}
	
	\caption{ Comparison of fluorescence data obtained using the BDM and the LPV model under multisine light intensity excitation.}\label{fig:mulisineresutls} 
\end{figure}

\noindent The comparison between the fluorescence results for the
BDM and LPV models is shown in Figure \ref{fig:mulisineresutls}.
The last period, with a fit of $R^{2}=0.97$, is presented in Figure
\ref{fig:mulisineresutls1}. To expose some frequency components,
Figures \ref{fig:mulisineresutls2}, \ref{fig:mulisineresutls3},
and \ref{fig:mulisineresutls4} illustrate the last $100$, $10$,
and $1$ seconds, respectively, to better relate the results to the light intensity excitation shown in Figure \ref{fig:multisinelight}.

%% file: Section_5_1_Conclusion.tex
This study has demonstrated the application of frequency-domain system
identification techniques for modeling the dynamic regulation of photosynthesis
under oscillating light, based on in-silico data generated from the
Basic DREAM Model (BDM). The Best Linear Approximation (BLA) method
successfully captured the local input--output dynamics, while the
subsequent Linear Parameter-Varying (LPV) formulation provided a compact
and scalable representation across a broad range of operating conditions
(i.e., in terms of the DC component of light intensity). Furthermore,
the use of multisine excitation signals proved to be an efficient
approach for characterizing the dynamical behavior of photosynthetic
regulation within short experimental time frames.

\noindent These results provide a first step toward data-driven, control-oriented modeling of photosynthesis regulation. Future research will focus
on validating the proposed framework using experimental plant data,
thereby bridging the gap between in-silico modeling and practical
applications in real-time phenotyping and optimized lighting strategies.

%% file: Section_6_1_Appendix01_BDM.tex
This Appendix gives a more detailed representation of the the BDM
framework presented in Section \ref{sec:Basic-DREAM-Model}. The vector
field $f(.)$ in Eq. \eqref{eq:x_dot} describes the evolution of
the state vector $\mathbf{x}(t)$, and is given by

\begin{equation}
f(\mathbf{x}(t),u(t),\mathbf{\Phi})=\left[\begin{array}{c}
\frac{v_{2}(t)-v_{1}(t)}{2}+v_{8}(t)\\
b_{H}\left(\frac{v_{PSII}(t)+v_{2}(t)}{N_{A}V_{L}}-\frac{14}{3}\frac{V_{S}}{V_{L}}v_{5}(t)-v_{7}(t)\right)\\
v_{3}(t)-v_{4}(t)\\
v_{5}(t)-v_{6}(t)\\
v_{PSI}(t)-v_{2}(t)
\end{array}\right],\label{eq:f_t}
\end{equation}

where the associated rates for $f(.)$ in Eq. \eqref{eq:f_t} are
given by 
\begin{subequations}
\label{eq:rates}

\begin{align}
v_{1}(t) & =k_{1}^{+}\textrm{RCII}_{CL}x_{1}(t)\ldots\nonumber \\
 & \ldots-k_{1}^{-}\left(1-\textrm{RCII}_{CL}\right)\left(\textrm{PQ}_{tot}-x_{1}(t)\right),\nonumber \\
 & =k_{1}^{-}\left(\textrm{RCII}_{CL}-1\right)\textrm{PQ}_{tot}\ldots,\nonumber \\
 & \ldots+\left(k_{1}^{+}\textrm{RCII}_{CL}+k_{1}^{-}\left(1-\textrm{RCII}_{CL}\right)\right)x_{1}(t)\label{eq:v1}\\
v_{2}(t) & =k_{2}^{+}\left(\textrm{PQ}_{tot}-x_{1}(t)\right)x_{5}(t)\ldots\nonumber \\
 & \ldots-k_{2}^{-}x_{1}(t)\left(1-x_{5}(t)\right),\nonumber \\
 & =k_{2}^{+}\textrm{PQ}_{tot}x_{5}(t)-k_{2}^{-}x_{1}(t)\ldots\nonumber \\
 & \ldots+\left(k_{2}^{-}-k_{2}^{+}\right)x_{1}(t)x_{5}(t),\label{eq:v2}\\
v_{3}(t) & =k_{3}\left(1-x_{3}(t)\right)\frac{1}{1+\left(\frac{K_{Q}}{x_{2}(t)}\right)^{n}},\label{eq:v3}\\
v_{4}(t) & =k_{4}x_{3}(t),\label{eq:v4}\\
v_{5}(t) & =k_{5}\left(A_{tot}-x_{4}(t)-\frac{x_{4}(t)}{\textrm{cEqP}}\left(\frac{H_{\textrm{stroma}}^{+}}{x_{2}(t)}\right)^{\frac{14}{3}}\right),\label{eq:v5}\\
v_{6}(t) & =k_{6}x_{4}(t),\label{eq:v6}\\
v_{7}(t) & =k_{7}\left(x_{2}(t)-H_{\textrm{stroma}}^{+}\right),\label{eq:v7}\\
v_{8}(t) & =v_{x}(t)=k_{8}\left(\textrm{PQ}_{tot}-x_{1}(t)\right),\label{eq:v8}\\
v_{PSII} & =n_{\textrm{PSII}}\sigma_{\textrm{II}}\left(1-\textrm{FQ}_{max}x_{3}(t)\right)\textrm{RCII}_{OP}L(t),\label{eq:vPSII}\\
v_{PSI} & =n_{\textrm{PSI}}\sigma_{\textrm{I}}\frac{L_{\nicefrac{1}{2}}L(t)}{(L_{\nicefrac{1}{2}}+L(t))}\left(1-x_{5}(t)\right).\label{eq:vPSI}
\end{align}
\end{subequations}

Note that based on \cite{Fuente2024PlantPhysio}, the following approximation holds: $\textrm{RCII}_{OP}+\textrm{RCII}_{CL}=1$, where we have 
\begin{equation}
\textrm{RCII}_{CL}=\frac{1}{1+\frac{k_{1}^{+}x_{1}(t)}{\left(1-\textrm{FQ}_{max}x_{3}(t)\right)L(t)+k_{1}^{-}\left(\textrm{PQ}_{tot}-x_{1}(t)\right)}}.\label{eq:RCII_CL}
\end{equation}

The vector field $g(.)$ in Eq. \eqref{eq:y} gives the trajectory
of the output vector $\mathbf{y}(t)$, and is written as follows:
\begin{equation}
g(\mathbf{x}(t),u(t),\mathbf{\Phi})=\left[\begin{array}{c}
g_{1}(\mathbf{x}(t),u(t),\mathbf{\Phi})\\
g_{2}(\mathbf{x}(t),u(t),\mathbf{\Phi})\\
g_{3}(\mathbf{x}(t),u(t),\mathbf{\Phi})
\end{array}\right],\label{eq:g_t}
\end{equation}
with 
\begin{subequations}
\label{eq:g_map} 
\begin{align}
g_{1}(\mathbf{x}(t),u(t),\mathbf{\Phi}) & =\left(1-\textrm{FQ}_{max}x_{3}(t)\right)\ldots\nonumber \\
 & \ldots\left(\frac{F_{0}}{F_{V}}+\textrm{RCII}_{CL}\right),\label{eq:g_1}\\
g_{2}(\mathbf{x}(t),u(t),\mathbf{\Phi}) & =\frac{\textrm{FQ}_{max}x_{3}(t)}{\left(1-\textrm{FQ}_{max}x_{3}(t)\right)},\label{eq:g_2}\\
g_{3}(\mathbf{x}(t),u(t),\mathbf{\Phi}) & =\frac{v_{PSII}}{4}\nonumber \\
 & =\frac{\left(1-\textrm{FQ}_{max}x_{3}(t)\right)\textrm{RCII}_{OP}L(t),}{4}.\label{eq:g_3}
\end{align}
\end{subequations}
 The following vector is constructed that consists of the BDM parameters
used in Eqs. \eqref{eq:f_t} -- \eqref{eq:g_3} \cite{Portilla2024ReportDeliverable}:
\begin{align}
\mathbf{\Phi} & =\left[\begin{array}{cccccccc}
kI & k_{1}^{+} & k_{1}^{-} & k_{2}^{+} & k_{2}^{-} & k_{3} & k_{4} & \cdots\end{array}\right.\nonumber \\
 & \left.\begin{array}{cccccc}
\cdots & k_{5} & k_{6} & k_{7} & k_{8} & \cdots\end{array}\right.\nonumber \\
 & \left.\begin{array}{cccccc}
\cdots & A_{tot} & b_{H} & \textrm{FQ}_{max} & n & \cdots\end{array}\right.\nonumber \\
 & \left.\begin{array}{cccc}
\cdots & \textrm{cEqP} & K_{Q} & \textrm{PQ}_{tot}\end{array}\right].\label{eq:Phi_parameter}
\end{align}
The BDM parameters for the vector fields are listed in \cite{Fuente2024PlantPhysio}.